\documentclass{appolb}
\pdfoutput=1

\usepackage{graphicx}
\usepackage[vcentermath]{youngtab}
\usepackage{amsmath,amsfonts}

\newcommand{\psiN}{\psi^{(N)}}

\newcommand{\la}{\langle}
\newcommand{\ra}{\rangle}

\newcommand{\braket}[1]{\mathinner{\left\langle{#1}\right\rangle}}

\newcommand{\sy}{\text{sym}}

\newcommand{\tr}{\text{true}}
\newcommand{\SU}{\text{SU}}
\newcommand{\1}{\mathbf{1}}

\newcommand{\N}{\mathbb{N}}

\DeclareMathOperator{\re}{Re}

\begin{document}
\pagestyle{plain}


\title{Eigenvalue density of Wilson loops in 2D $\SU(N)$ YM at large $N$%
\thanks{Presented at Cracow School of Theoretical Physics, Zakopane, 31.5.-10.6.2009}%
}
\author{Robert Lohmayer
\address{Institute for Theoretical Physics, University of
  Regensburg,\\ 93040 Regensburg, Germany}
}

\maketitle
\begin{abstract}
The eigenvalue density of a Wilson loop matrix $W$ associated with a simple loop in two-dimensional Euclidean $\SU(N)$ Yang-Mills theory
  undergoes a phase transition at a critical size in the infinite-$N$ limit. The averages of $\det(z-W)^{-1}$ and $\det(1+uW)/(1-vW)$ at finite $N$ lead to two different smoothed out expressions. It is shown by a saddle-point analysis that both functions tend to the known singular result at infinite $N$.  
\end{abstract}
  
\section{Introduction}
In two Euclidean dimensions the eigenvalue distribution of the
$\SU(N)$ Wilson matrix associated with a non-selfintersecting loop
undergoes a phase transition in the infinite-$N$ limit as the loop is
dilated \cite{Durhuus:1980nb}.  This phase transition has universal
properties shared across dimensions and across analog two-dimensional
models \cite{Narayanan:2007dv,Narayanan:2008he}. Thus, a detailed
understanding of the transition region in 2D is of relevance to
crossovers from weakly to strongly interacting regimes in a wide class
of models based on doubly indexed dynamical variables with symmetry
$\SU(N)$.
Building upon previous work
\cite{Neuberger:2008mk,Neuberger:2008ti,Blaizot:2008nc}, new results in this context have been obtained in \cite{Lohmayer:2009aw}. Some of these results are presented here.

We are focusing on the eigenvalues of the Wilson loop. The associated
observables are two different functions $\rho_N^\tr (\theta)$, $\rho_N^\sy (\theta)$  of an angular variable $\theta$. At infinite
$N$ the two functions have identical limits: $\rho_\infty^{\tr}(\theta)=\rho_\infty^{\sy}(\theta)\equiv\rho_\infty(\theta)$.

For a specific critical scale, the
nonnegative function $\rho_\infty (\theta )$ exhibits a transition at
which a gap centered at $\theta=\pm \pi$, present for small loops,
just closes. This transition was discovered by Durhuus and Olesen in
1981~\cite{Durhuus:1980nb}. 

\section{Eigenvalue densities}
\label{sec:evdens}

The probability density for the Wilson loop matrix $W$ is given by the heat kernel (see for
example~\cite{Gross:1993hu} and original references therein)
\begin{equation}
  \label{eq:weight}
  {\cal P}_N (W,t) = \sum_r d_r \chi_r (W) e^{-\frac{t}{2N} C_2 (r)}
\end{equation}
with $t=\lambda {\cal A}$, where $\lambda=g^2N$ is the standard 't Hooft coupling and $\cal A$ denotes the area enclosed by the loop. The sum over $r$ is over all distinct
irreducible representations of $\SU(N)$ with $d_r$ denoting the
dimension of $r$ and $C_2 (r)$ denoting the value of the quadratic
Casimir on $r$. $\chi_r (W)$ is the character of $W$ in the
representation $r$ and is normalized by $\chi_r (\1) = d_r$.  Averages
over $W$ at fixed $t$ are given by
\begin{equation}
  \langle {\cal O}(W)\rangle = \int dW {\cal P}_N (W,t )\mathcal{O}(W)\,,
\end{equation}
where $dW$ is the Haar measure on $\SU(N)$ normalized by $\int dW =1$.
Note that we have $\int dW {\cal P}_N(W,t)=1$.
Any class function can be averaged when expanded in characters using
character orthogonality.  

Because in the sum over $r$ in \eqref{eq:weight} each representation
is accompanied by its complex conjugate representation, it is easy to
see that
\begin{equation} 
  \langle {\cal O}(W)\rangle = \langle {\cal O}(W^\dagger)\rangle =
  \langle {\cal O}(W^\ast)\rangle\, , 
\label{symmetry}
\end{equation}
implying identities relating $\langle \det(z-W)\rangle$, $\langle
\det(z-W)^{-1}\rangle$, and $\langle \det(1+uW)/(1-vW)\rangle$ to the
same objects with $z\to 1/z$, $z\to z^*$, $u,v\to 1/u ,1/v$, and
$u,v\to u^*, v^*$, respectively.

The density functions $\rho^\sy_N$ and $\rho^\tr_N$ are obtained from
\begin{align}
G^{\tr}_N (z) &=\frac{1}{N} \braket{ \Tr\frac{1}{z-W} }= \frac{1}{N} \frac{\partial}{\partial z} \langle \log 
\det (z-W)\rangle,\\
G^{\sy}_N (z) &= -\frac{1}{N} \frac{\partial}{\partial z} \log \langle \det \left (z-W\right )^{-1} \rangle\label{Gsy}
\end{align}
through ($\ell=\tr,\sy$)
\begin{align}
\rho^{\ell}_N(\theta)&=2 \lim_{\epsilon\to0^+} \re\left[e^{i\theta+\epsilon}G^{\ell}_N(e^{i\theta+\epsilon})\right]-1\,.
\label{eq:rho}
\end{align}
The $\rho^\ell_N$ are real on the unit circle parametrized by the
angle $|\theta|\le\pi$, even under $\theta\to-\theta$, and depend on
the size of the loop. Both are positive distributions in
$\theta$, normalized by
\begin{equation}
  \int_{-\pi}^\pi \frac{d\theta}{2\pi} \rho^\ell_N (\theta )=1\,.
\end{equation}

Only $\rho_N^\tr$ has a natural interpretation at finite $N$, it literally is the eigenvalue density. If the eigenvalues
of $W$ are $e^{i\alpha_j}$ with $j=0,1,\ldots,N-1$, it is given by \cite{Lohmayer:2009aw}
\begin{align}
  \rho^{\tr}_N(\theta)
  =\frac{2\pi}{N}\sum_j\langle \delta_{2\pi}(\theta-\alpha_j(W))\rangle
  =\frac{2\pi}{N}\langle \Tr\delta_{2\pi}(\theta+i\log(W))\rangle\,.
\end{align}
$\rho^{\tr}_N$ determines $\langle \Tr f(W)\rangle $ for any function $f$.

The density $\rho_N^\sy$ is determined by the averages of the characters of $W$ in all totally symmetric representations of $\SU(N)$. This function is of interest mainly because it obeys simple partial integro/differential equations which are exactly integrable \cite{Neuberger:2008ti}.

\begin{figure}
  \includegraphics[width=0.44\textwidth]{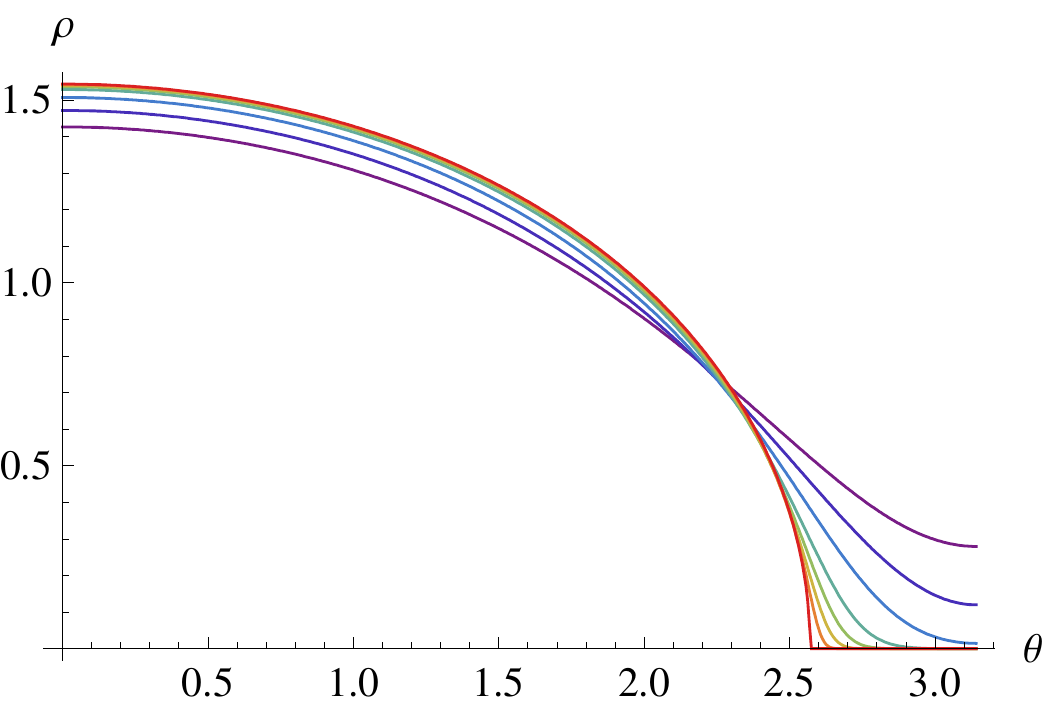}\hfill    
  \includegraphics[width=0.44\textwidth]{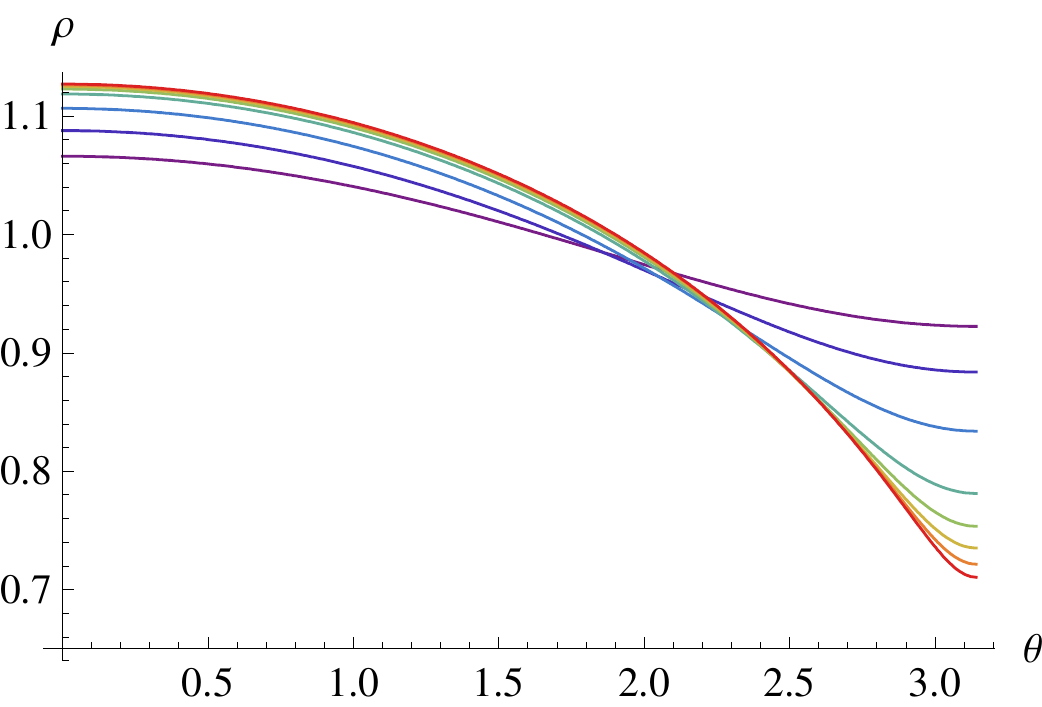}\hfill
  \vspace{-2cm}    
  \includegraphics[width=0.09\textwidth]{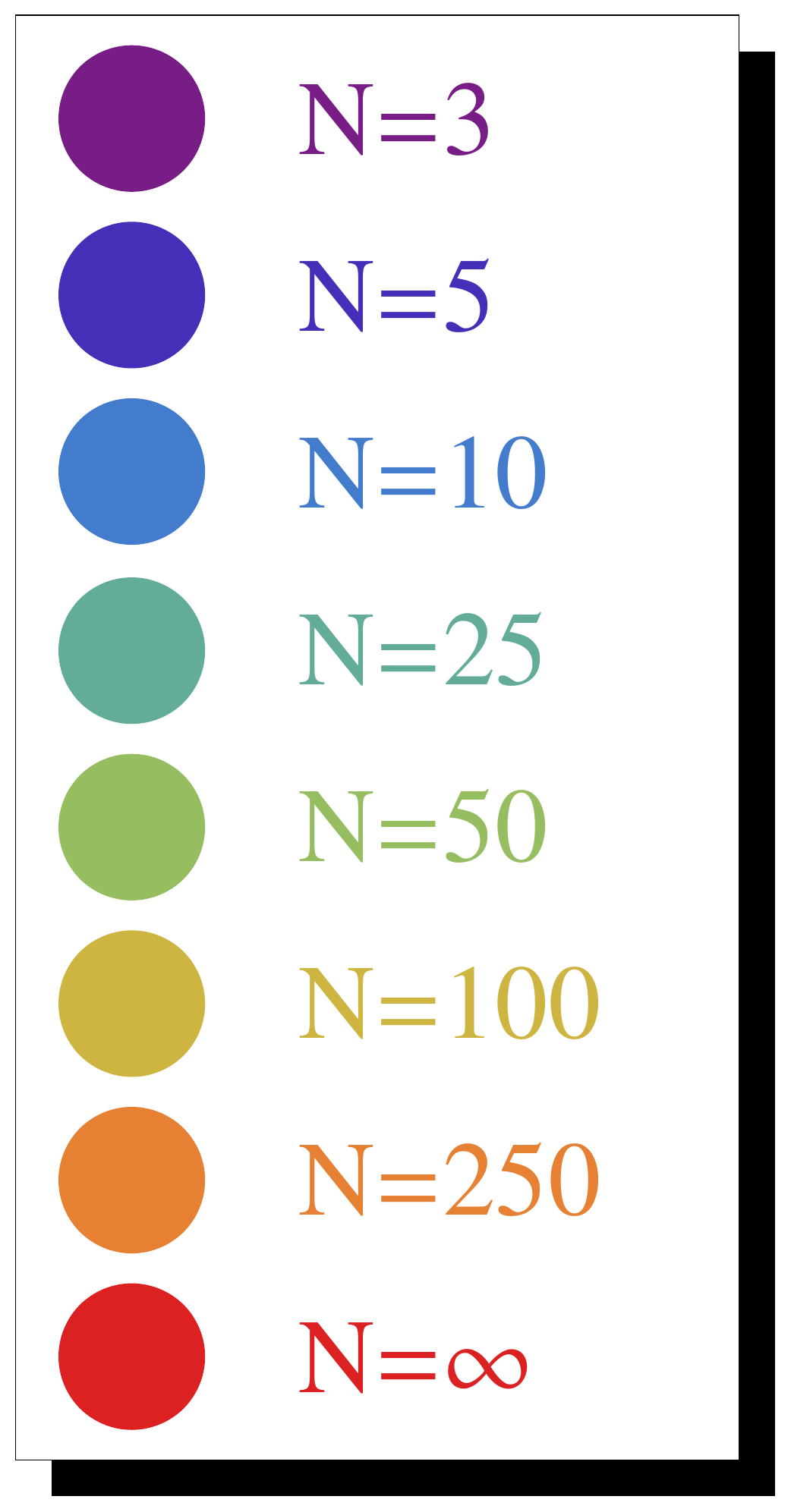} 
  \vspace{2cm}
  \caption{Plots of $\rho_N^\sy(\theta,T)$ for $T=2$ (left), $T=5$
    (right), and $N=3,5,10,25,50,100,250$ together with
    $\rho_\infty(\theta,T)$.}
  \label{figRhoSymm}
 \end{figure}

$\rho^{\sy}_N$ has an explicit form in terms of
rapidly converging infinite sums \cite{Neuberger:2008ti}, which can be evaluated numerically for arbitrary $N$ to any desired precision. In Fig.~\ref{figRhoSymm} we show how
$\rho_N^\sy(\theta)$ approaches the infinite-$N$ result
$\rho_\infty(\theta)$ of Durhuus and Olesen~\cite{Durhuus:1980nb}. 
$\rho^{\sy}_N$ is monotonic on each of the segments $(-\pi,0)$ and $(0,\pi)$ with the maximum at $\theta=0$ and the minimum at $\theta=\pm\pi$. 
In addition to these numerical results, it would be useful to compute
analytically the asymptotic expansion of $\rho^{\sy}_N (\theta)$ in
$1/N$ (cf. section \ref{Sec:sym}).

It turns out that the appropriate area variable for $\rho_N^\sy$ is not $t=\lambda {\cal A}$ but $T=t(1-1/N)$ \cite{Neuberger:2008ti}. When $\rho^{\sy}_N (\theta,T)$ is compared to $\rho^{\tr}_N
(\theta, t)$, the $1/N$ correction in $t$ relative to $T$ has to be
taken into account (so far, the size dependence of the $\rho_N^\ell$ has been suppressed).

The infinite-$N$ critical point is at $T=4$. For $T>4$, $\rho^{\sy}_N
(\theta ,T)$ approaches $\rho_\infty (\theta,T)$ by power corrections
in $1/N$ \cite{Neuberger:2008ti}.  For $T<4$, $\rho_\infty (\theta,T)$
is zero for $|\theta|>\theta_c(T)$, where $0<\theta_c(T)<\pi$ and
$\theta_c(4)=\pi$ (cf. Fig.~\ref{figRhoSymm}).  In this interval $\rho^{\sy}_N (\theta ,T)$
approaches zero by corrections that are exponentially suppressed in
$N$.

\begin{figure}
  \begin{tabular}{l@{\hspace*{5mm}}r}
    \includegraphics[width=0.46\textwidth]{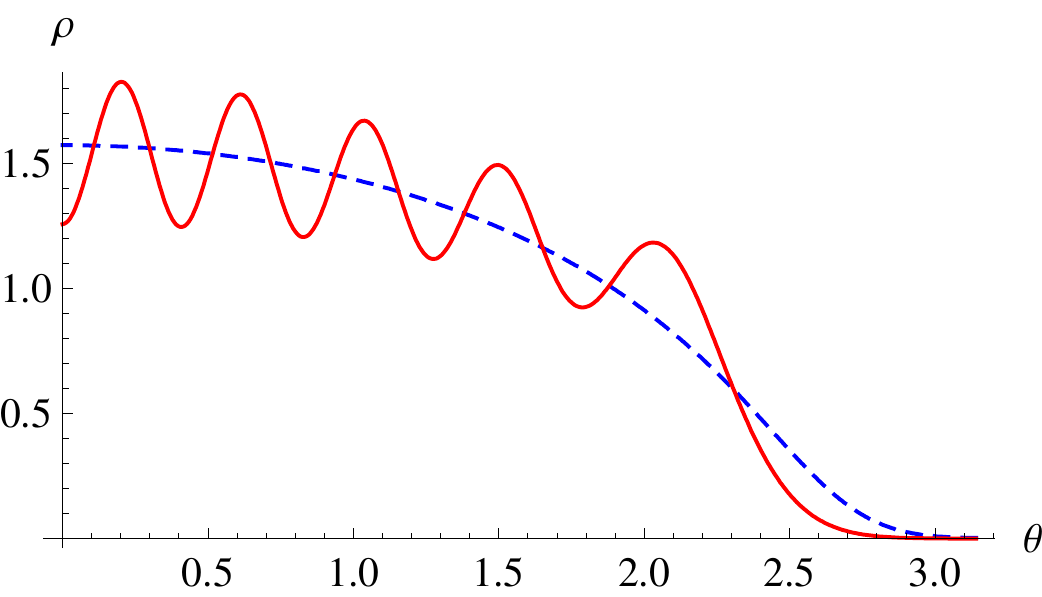} &
    \includegraphics[width=0.46\textwidth]{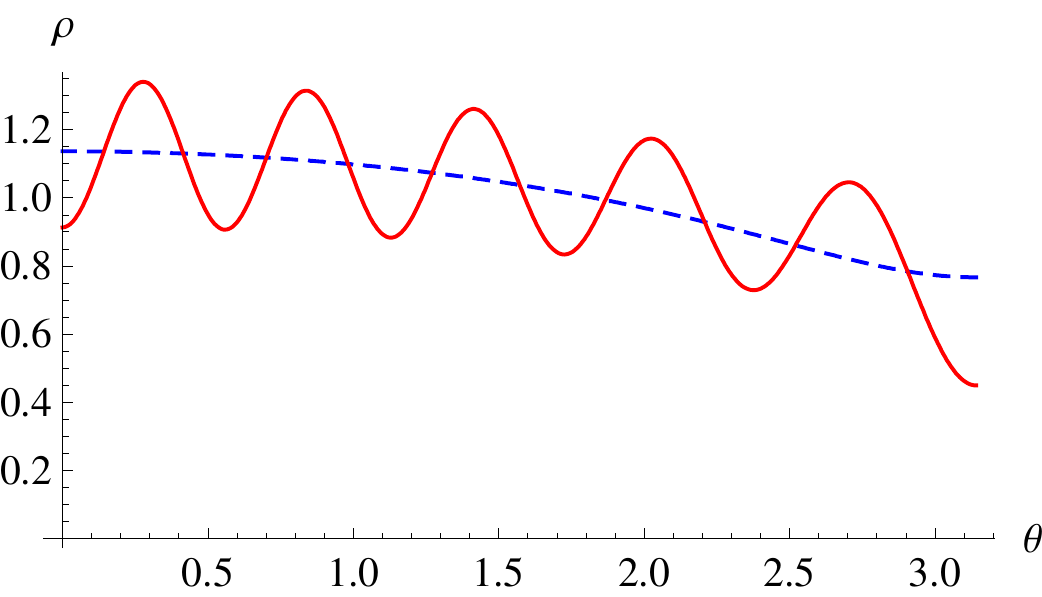} \\
    \includegraphics[width=0.46\textwidth]{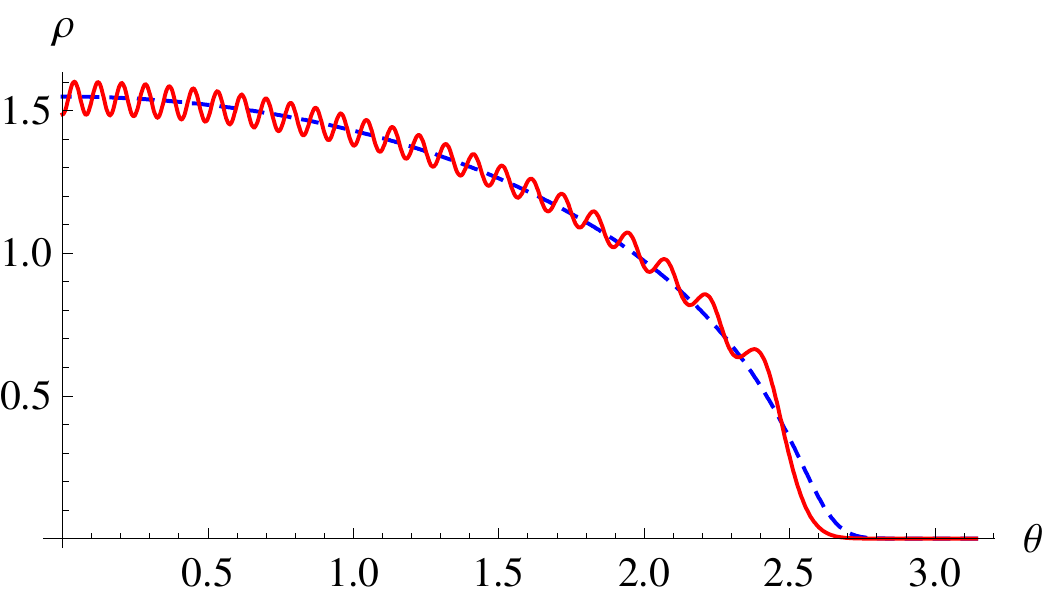} &
    \includegraphics[width=0.46\textwidth]{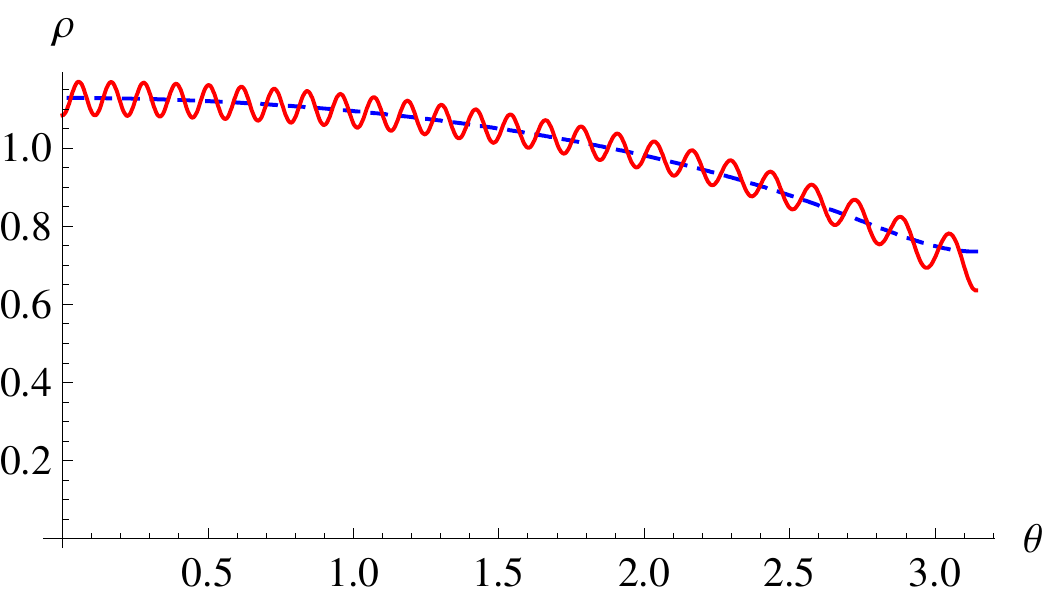} 
  \end{tabular}
  \caption{Plots of the densities $\rho_N^\tr(\theta,t)$ (red, solid)
    and $\rho_N^\sy(\theta,T)$ (blue, dashed) for $t=2$ (left) and
    $t=5$ (right), $N=10$ (top), and $N=50$ (bottom).}
  \label{figRhoTrueSymm}
\end{figure}

The true eigenvalue density $\rho_N^\tr(\theta,t)$ has $N$ peaks (in the interval $[-\pi,\pi]$) and oscillates around $\rho_N^\sy(\theta,T)$ (cf. \cite{Lohmayer:2009aw}). Plots of  $\rho_N^\tr(\theta,t)$ and $\rho_N^\sy(\theta,T)$ are shown in Fig.~\ref{figRhoTrueSymm} for $t=2$ and $t=5$. 

\section{Integral representations}

The density $\rho_N^\tr$ can be obtained from the expectation value of \cite{Lohmayer:2009aw}
\begin{align}
R(u,v,W)=\frac{\det(1+uW)}{\det(1-vW)}=\sum_{p=0}^N\sum_{q=0}^\infty u^p v^q \chi_p^A (W) \chi_q^S (W)
\label{charexp}
\end{align}
with $|v|<1$. $\chi_p^A(W)$ (resp. $\chi_q^S(W)$) denotes the character of $W$ in a
totally antisymmetric (resp. symmetric) representation. When we set $u=-v+\epsilon$ and expand to linear order in~$\epsilon$, the LHS reads
\begin{align}
R(-v+\epsilon, v ,W)=1-\epsilon \Tr\frac{1}{v-W^\dagger}\,.
\end{align}
After decomposing the tensor product $p^A\otimes q^S$ into irreducible representations, we obtain for the expectation value of the trace (due to character orthogonality) 
\begin{align}
\bar R(v)\equiv \braket{ \Tr\frac{1}{v-W^\dagger } } = -
  \sum_{p=0}^{N-1}\sum_{q=0}^\infty (-1)^p v^{p+q} e^{-\frac{t}{2N}
    C(p,q)} d(p,q)\,,
\label{barR}
\end{align}
where $C(p,q)$ and $d(p,q)$ denote the value of the quadratic Casimir and the dimension of the irreducible representation identified by the Young diagram
\begin{align}
  \qquad\qquad\young(\hfil12\hfil\hfil q,1,\hfil,\hfil,p)
\end{align}
and are given by \cite{Perelomov:1965ab}
\begin{align}
    C(p,q)&=(p+q+1)\left(N-\frac{p+q+1}{N}+q-p\right)\label{Cpq}\,,\\
   d(p,q) &=d^A (p) d^S (q) \frac{(N-p)(N+q)}{N} \frac{1}{p+q+1}\,,\\
  d^A (p) &= {N \choose p}\,,\qquad d^S(q) = {N+q-1\choose q}.
\end{align}
We can exactly calculate sums of the form (with $|z|<1$)
\begin{align}
\sum_{p=0}^{N-1} u^p d^A(p) (N-p) =
    N(1+u)^{N-1}\,,\quad \sum_{q=0}^\infty z^q d^S (q) (N+q) =
    \frac{N}{(1-z)^{N+1}}\,.
\end{align}
To factorize the sums over $p$ and $q$ in \eqref{barR}, we first write
\begin{align}
\frac{1}{p+q+1}=\int_0^1 d\rho \rho^{p+q}\,.
\end{align}
The $t$-dependent weight factor is the exponent of a bilinear form in
$p$ and $q$ (given by \eqref{Cpq}). By a Hubbard-Stratonovich transformation the dependence on $p$ and $q$ can be made linear. Performing the (independent) sums over $p$, $q$ then leads to \cite{Lohmayer:2009aw}
\begin{align}
\bar R(v)&=-\frac{N^2}{t}e^{-\frac t2}\int\!\!\int_{-\infty}^\infty
  \frac{dxdy}{2\pi}\int_0^1d\rho\, 
  e^{-\frac N{2t}(x^2+y^2)+\frac1{2t}(x+iy)^2-\frac12(x-iy)} \nonumber\\
  &\quad\times
  \frac{\left[1-v\rho e^{-x-t/2}\right]^{N-1}}{\left[1-v\rho e^{iy-t/2}\right]^{N+1}}
  \label{Rint}
\end{align}
(valid for $|v|<1$). Now the entire dependence on $N$ is explicit. The infinite-$N$ limit of $\rho_N^\tr$ can be obtained from this integral representation by using a saddle-point approximation for the integrals over $x$ and $y$ (cf. Sec.~\ref{Sec:SaddleTrue}).

\section{Asymptotic expansion of  $\rho^\sy_N$}
\label{Sec:sym}

An integral representation for $\psiN(z)=\la\det(z-W)^{-1}\ra$, which determines $\rho^\sy_N$, is obtained in a similar manner \cite{Neuberger:2008ti}. In this case, only a single integral is needed (valid for $|z|>1$),
\begin{align}
\psiN(z)=e^{\frac{NT}{8} } \sqrt{\frac{N}{2\pi T}}
  \int_{-\infty}^\infty du \, e^{-\frac{N}{2T}u^2} \left ( z
    e^{-i\frac{u}{2}} - e^{i\frac{u}{2}}\right )^{-N}\,.
    \label{psint}
\end{align}
The aim of this section is to construct an asymptotic expansion of
$\rho_N^\sy(\theta,T)$ in powers of $1/N$ (cf. \cite{Lohmayer:2009aw}).  To this end we perform a
saddle-point analysis of the integral in \eqref{psint}, from which
$\rho_N^\sy$ can be obtained via \eqref{Gsy} and \eqref{eq:rho}. 

\subsection{Saddle-point analysis}
\label{sec:spa}

For $|z|=1$ the integrand of \eqref{psint} has singularities on the
real-$u$ axis.  We therefore set $z=e^{\epsilon+i\theta}$, where
$\epsilon >0$ ensures that $|z|>1$ but will later be taken to zero.
The integrand of \eqref{psint} can be written as $\exp(-Nf(u))$ with
\begin{align}
  f(u)=\frac{u^2}{2T}+\log\left(ze^{-i\frac u2}-e^{i\frac u2}\right)\,.
\end{align}
We now look for saddle points of the integrand in the complex-$u$
plane, which we label by $\bar u=iTU(\theta,T)$, where
$U(\theta,T)=U_r (\theta,T) + i U_i (\theta, T)$ is a complex-valued
function of $\theta$ and $T$.  The saddle-point equation turns out to
be
\begin{equation}
  e^{-TU(\theta,T)}\frac{U(\theta, T) + 1/2}{U(\theta, T) -1/2}
  =e^{\epsilon+i\theta}\,.
  \label{contour}
\end{equation}
For $\epsilon=0$, this is equation (5.49) in~\cite{Lohmayer:2008bd}
and is related to the inviscid complex Burgers equation via equation
(5.44) there. In the present notation, the latter equation has the
form
\begin{equation} 
  \frac{\partial U}{\partial T}+iU\frac{\partial
    U}{\partial \theta} =0\,.
\end{equation}
Taking the absolute value of \eqref{contour} leads to the equation
\begin{align}
  \label{eq:uiur}
  U_i^2 &= U_r\coth (T U_r+\epsilon)-U_r^2 -\frac{1}{4}\,.
\end{align}
For $\epsilon=0$, this equation has been investigated previously
in~\cite{Lohmayer:2008bd}.  However, here we keep $\epsilon>0$ for the
time being.  The singularities of the integrand of \eqref{psint} then
all have $U_r<0$.  Equation~\eqref{eq:uiur} describes one or more
curves in the complex-$U$ plane on which the saddle points have to lie
(for a given value of $\theta$, the saddles are isolated points on
these curves). 

\begin{figure}
  \includegraphics[width=.32\textwidth]{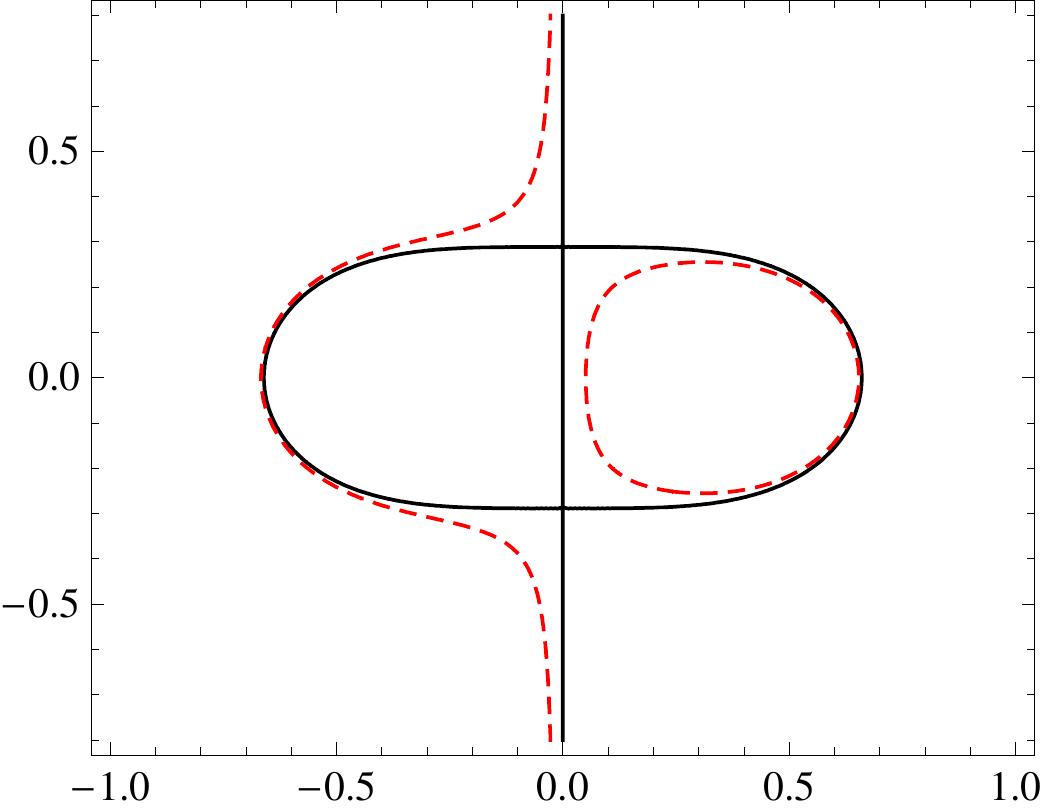}\hfill
  \includegraphics[width=.32\textwidth]{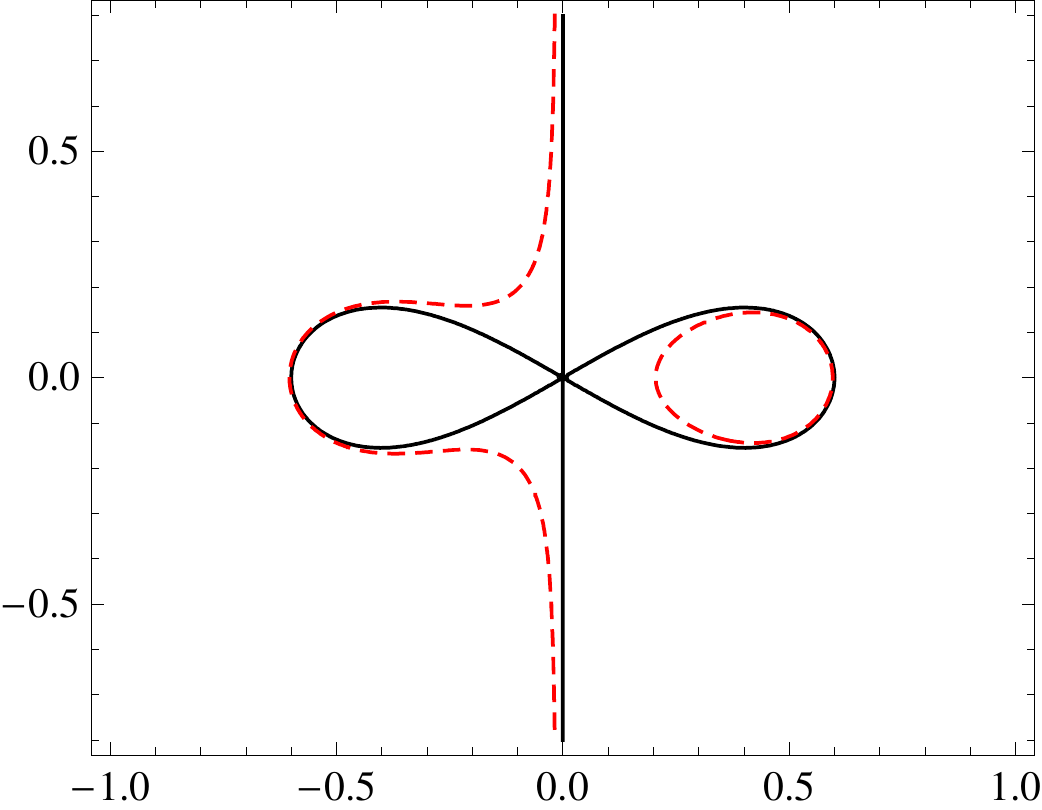}\hfill
  \includegraphics[width=.32\textwidth]{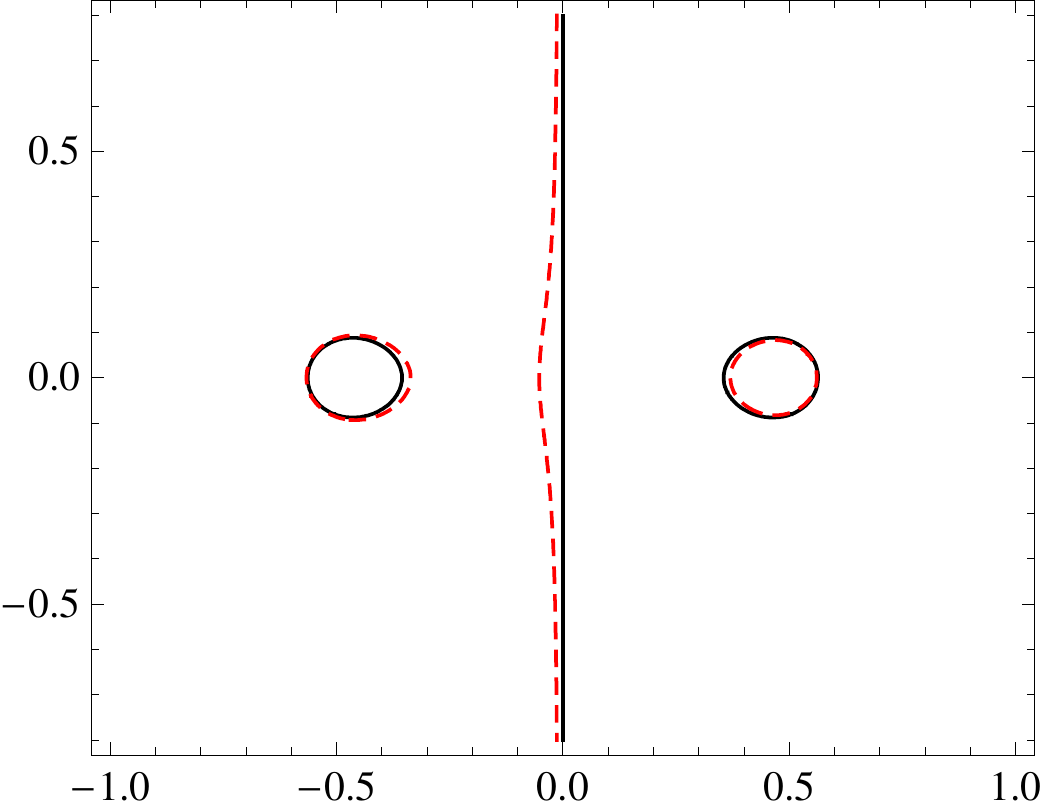}
  \caption{Examples of the contours in the complex-$U$ plane described by equation ({\protect\ref{eq:uiur}}) 
  for $T=3$ (left), $T=4$ (middle), and $T=5$ (right). The red dashed curves are for small $\epsilon>0$, while the black solid curves are for $\epsilon=0$.  For our saddle-point
    analysis we keep $\epsilon>0$.}
   \label{fig:curves}
\end{figure}

\begin{figure}
\begin{minipage}{0.45\textwidth}
\begin{center}
  \includegraphics[width=.9\textwidth]{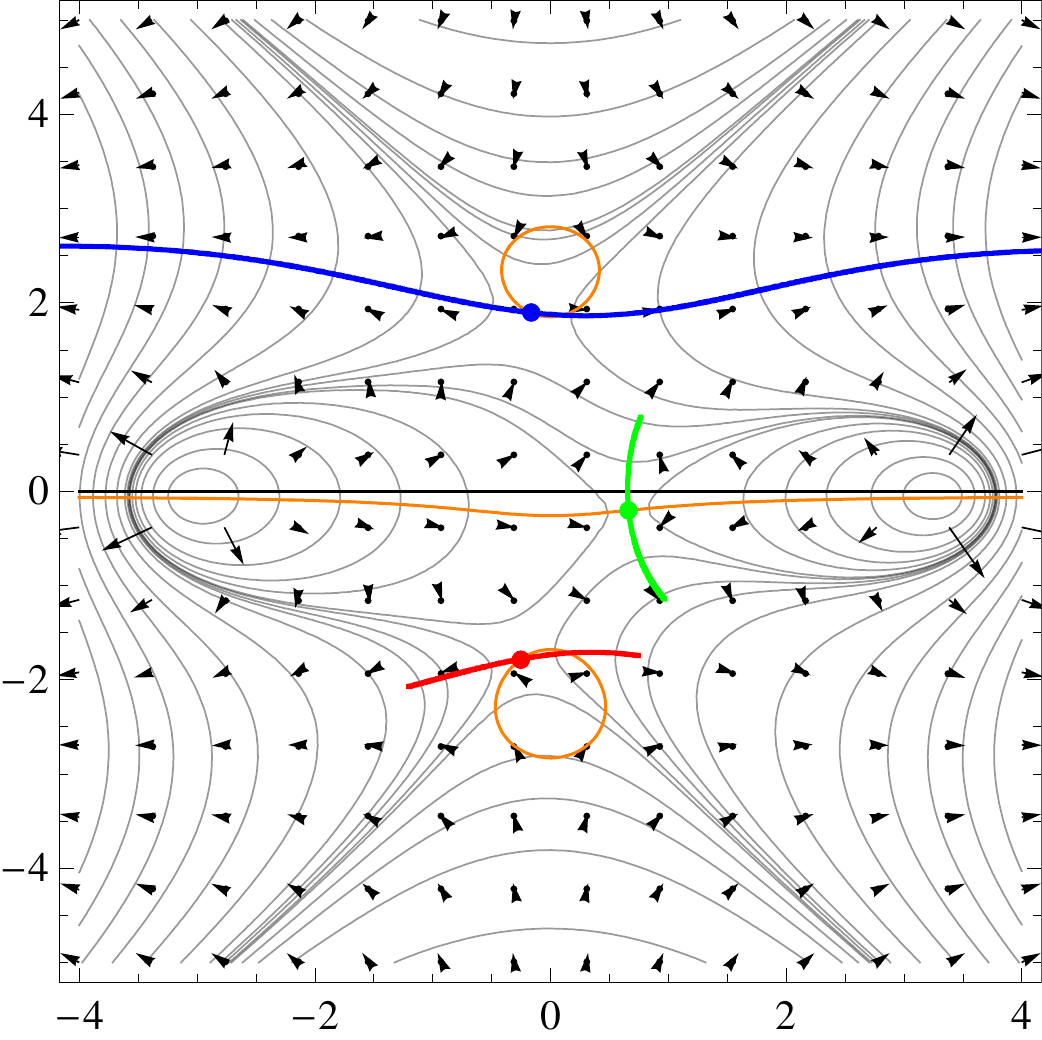}
\end{center}
\end{minipage}
\begin{minipage}{0.52\textwidth}
	\caption{
Example for the location of the saddle points and the deformation of the integration contour in the complex-$u$ plane for $T=5$ and $\theta=3$ (with small $\epsilon>0$). The thin solid lines are lines of constant $\re f(u)$, the arrows point in the direction of increasing $\re f(u)$. On each of the closed orange curves there is one saddle point (red dot and blue dot), and on the open orange curve there are infinitely many saddle points, but only one of them in the region shown in the plot (green dot). The thick blue curve is the integration path along the direction of steepest descent through the relevant saddle point.}
\label{figuCont}
\end{minipage}
\end{figure}

In Fig.~\ref{fig:curves} we show typical examples for these curves
for $T<4$, $T=4$, and $T>4$, where $\epsilon$ has been chosen
sufficiently close to zero.  (The closed contours always enclose the
points $U=1/2$ or $U=-1/2$.  For $T>4$ and larger $\epsilon$, the
closed contour in the left half-plane would be missing, but right now
we are not concerned with this since we are only interested in the
limit $\epsilon\to0^+$.)  Analyzing \eqref{contour} numerically we
find, for all values of $T$, that for a given value of $\theta$ there
is always one (and only one) saddle point on the closed contour in the
right half-plane, i.e., with $U_r>0$.  Note that we are showing the
complex-$U$ plane, in which the original integration contour
corresponds to the imaginary axis.  The integration contour can be
smoothly deformed to go through the (single) saddle point in the right
half-plane along a path of steepest descent.  No singularities are
crossed since they all have $U_r<0$.  There are also saddle points on
the contour(s) in the left half-plane (in fact, there are infinitely
many on the open contour), but these need not be considered. 
Figure~\ref{figuCont} shows an example for the location of the saddle 
points and the deformation of the integration contour in the complex-$u$ 
plane.

Once the integration contour has been deformed to go through the
saddle point, we can safely take the limit $\epsilon\to0^+$.
Parametrizing the contour in the vicinity of the saddle point by
$u=\bar u+xe^{i\beta}$, where $x$ is the new integration variable
corresponding to the fluctuations around the saddle and $\beta$ is the
angle which the path of steepest descent makes with the real-$u$ axis,
$\psiN (e^{i\theta},T)$ is given, up to exponentially small
corrections in $N$, by
\begin{align}
  \psiN (e^{i\theta},T) &= \frac1{2^N} \sqrt{\frac{N}{2\pi
      T}}\, e^{\frac{NT}{8}-i\frac{N\theta}{2}+i\beta}
  \int_{-\infty}^\infty
  dx \, e^{-Ng(x)}\,,\\
  g(x)&=\frac1{2T}\bigl(xe^{i\beta}+iTU(\theta,T)\bigr)^2
  +\log\sinh\frac{i\theta-ixe^{i\beta}+TU(\theta,T)}{2}\,.
\end{align}
We can now expand $g(x)$ in $x$.  The linear order vanishes by
construction.  The second order gives a Gaussian integral over $x$,
resulting in
\begin{align}
  \psiN (e^{i\theta},T)&\approx 
  e^{\frac{NT}{8}+\frac{N T U^2 (\theta, T)}2}
  \frac{\left[e^{-i\theta}(1/4-U^2(\theta,T))\right]^{N/2}}
  {\sqrt{1-T(1/4-U^2(\theta,T))}} \,.
  \label{eq:spgauss}
\end{align}
Note that the factor $e^{-i\theta}$ cannot be pulled out of the term
in square brackets because periodicity in $\theta$ would be lost.

There is a potential complication.  In principle, $g''(0)$ and
therefore the denominator in \eqref{eq:spgauss} could be zero, which
would mean that the integral over $x$ cannot be performed in Gaussian
approximation.  For $T>4$, it is straightforward to show that $g''(0)$
is never zero.  For $T\le4$, one can use \eqref{contour} to show that
$g''(0)=0$ only for the saddle points corresponding to the two angles
$\theta=\pm\theta_c(T)$ at which $\rho_\infty(\theta,T)$ becomes zero
(see section~\ref{sec:evdens}).  This means that for
$|\theta|=\theta_c(T)$ the asymptotic expansion in $1/N$ diverges, and
that it converges ever more slowly as $|\theta|\to\theta_c$ from
below.

Note that for $T<4$ and $\theta_c(T)\le|\theta|\le\pi$ the function
$\rho^\sy_N(\theta,T)$ is exponentially suppressed in $N$.  The study of
the large-$N$ asymptotic behavior in this region requires more work.

\subsection{Leading-order result}

Equation~\eqref{eq:spgauss} is the leading order in the $1/N$
expansion of $\psiN (e^{i\theta},T)$.  We now show that it
leads to $\rho_N^\sy(\theta,T)\to\rho_\infty(\theta,T)$ as
$N\to\infty$.  We first write \eqref{eq:spgauss} in the form
\begin{align}
  \frac1N\log\psiN(e^{i\theta},T)=\frac T8-f(\bar u)+\mathcal
  O(1/N)\,.
\end{align}
Note that in this order we do not need the denominator in
\eqref{eq:spgauss}, which corresponds to $f''(\bar u)$ (or $g''(0)$).
Using $\bar u=iTU$ this leads to
\begin{align}
G^\sy_N(z)=\frac{1}{z-e^{-TU}}+\mathcal O(1/N)=\frac1z\left(U+\frac{1}{2}\right)+\mathcal O(1/N)  
\end{align}
where in the last step we have used the saddle-point equation
\eqref{contour}.  Equation \eqref{eq:rho} then gives
\begin{align}
  \lim_{N\to\infty}\rho_N^\sy(\theta,T)=2\re U(\theta,T)\,,
\end{align}
which equals $\rho_\infty(\theta,T)$ of DO
\cite{Durhuus:1980nb,Janik:2004tw} since $U(\theta,T)$ satisfies
\eqref{contour} (which leads to \eqref{SaddleLambda} below with
$\lambda=U-1/2$ and $v=1/z$).

\subsection{$1/N$ correction to $\rho_\infty$}
\label{sec:spa1N}

Higher-order terms in the $1/N$ expansion of $\psiN
(e^{i\theta},T)$ can be obtained in the standard way by considering
higher powers of $x$ in the expansion of $g(x)$, resulting in
integrals of the type $\int_{-\infty}^\infty dx\,x^{2n}e^{-g''(0)x^2/2}$ with $n\in\N$.  However, if we are
only interested in the $1/N$ correction to $\rho_\infty(\theta,T)$ the
result \eqref{eq:spgauss} is already sufficient ($1/N$ corrections to
this result would give $1/N^2$ corrections to
$\rho_\infty(\theta,T)$).  Therefore we now write
\begin{align} 
  \frac1N\log\psiN(e^{i\theta},T)=\frac T8-f(\bar u)
  -\frac1{2N}\log[Tf''(\bar u)]+\mathcal O(1/N^2)\,,
\end{align}
which leads to
\begin{align}
  \label{eq:rhosym1N}
  \rho_N^\sy(\theta,T)=2\re \left[U \left(1+\frac 1N
      \frac{T(1/4-U^2)}{[1-T(1/4-U^2)]^2}\right)\right]+\mathcal
  O(1/N^2) \,.
\end{align}
Note that for $T\le4$ and $|\theta|\to\theta_c(T)$ (from below) the
denominator of the $1/N$ term approaches zero, which corresponds to
the complication discussed in section~\ref{sec:spa}.  Note also that
for $T\le4$ and $|\theta|>\theta_c$ the saddle point $U(\theta,T)$ is
purely imaginary so that both the leading order and the $1/N$ term are
zero.  This confirms that the above saddle-point analysis is not the
right tool to compute finite-$N$ effects in this region.

\begin{figure}
\includegraphics[width=0.43\textwidth]{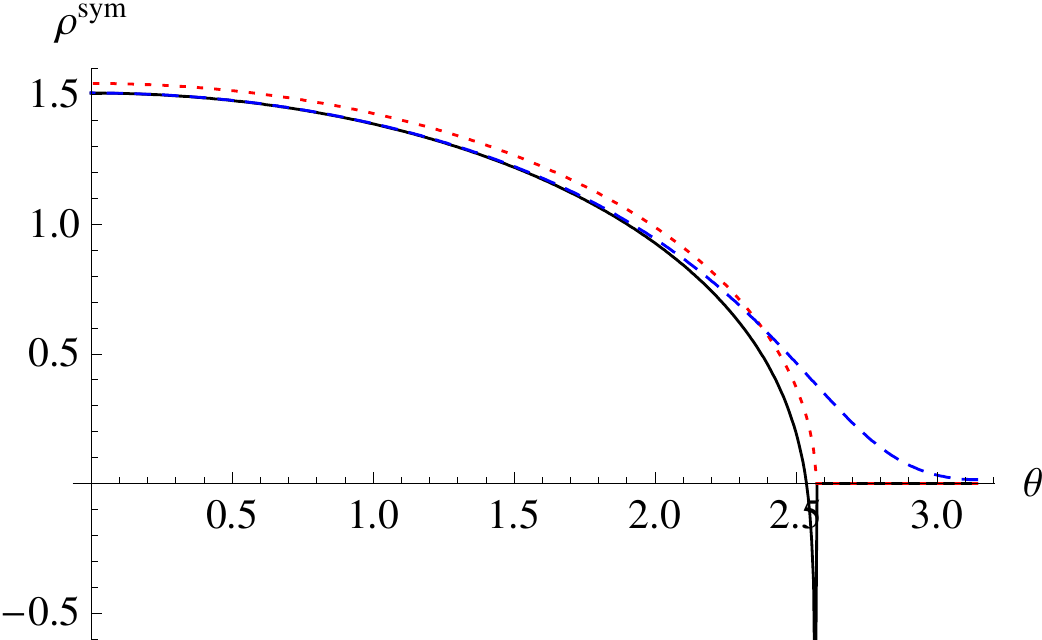}\hfill
  \includegraphics[width=0.43\textwidth]{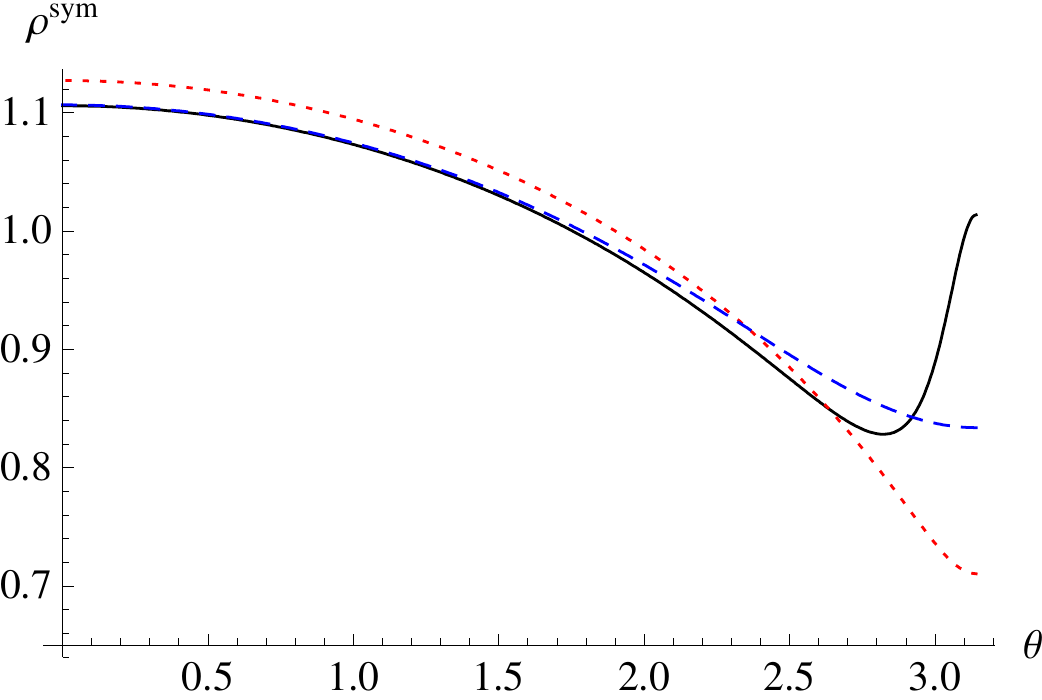}
  \caption{Examples for the $1/N$ corrections to
    $\rho_\infty(\theta,T)$ for $N=10$, $T=2$ (left), and $T=5$
    (right).  Shown are the exact result for $\rho_N^\sy(\theta,T)$
    (blue dashed curve), the infinite-$N$ result
    $\rho_\infty(\theta,T)$ (red dotted curve), and the asymptotic
    expansion of $\rho_N^\sy(\theta,T)$ up to order $\mathcal O(1/N)$ (black solid curve).  We observe that the asymptotic expansion converges rapidly for small $|\theta|$ and
    more slowly for larger $|\theta|$.}
  \label{fig:1N} 
\end{figure}

In Fig.~\ref{fig:1N} we show examples for the $1/N$ corrections to
$\rho_\infty(\theta,T)$ for $N=10$ and $T=2$ and $5$. 

\section{Saddle-point analysis for $\rho^\tr_N$}
\label{Sec:SaddleTrue}

\begin{figure}
  \includegraphics[width=0.3\textwidth]{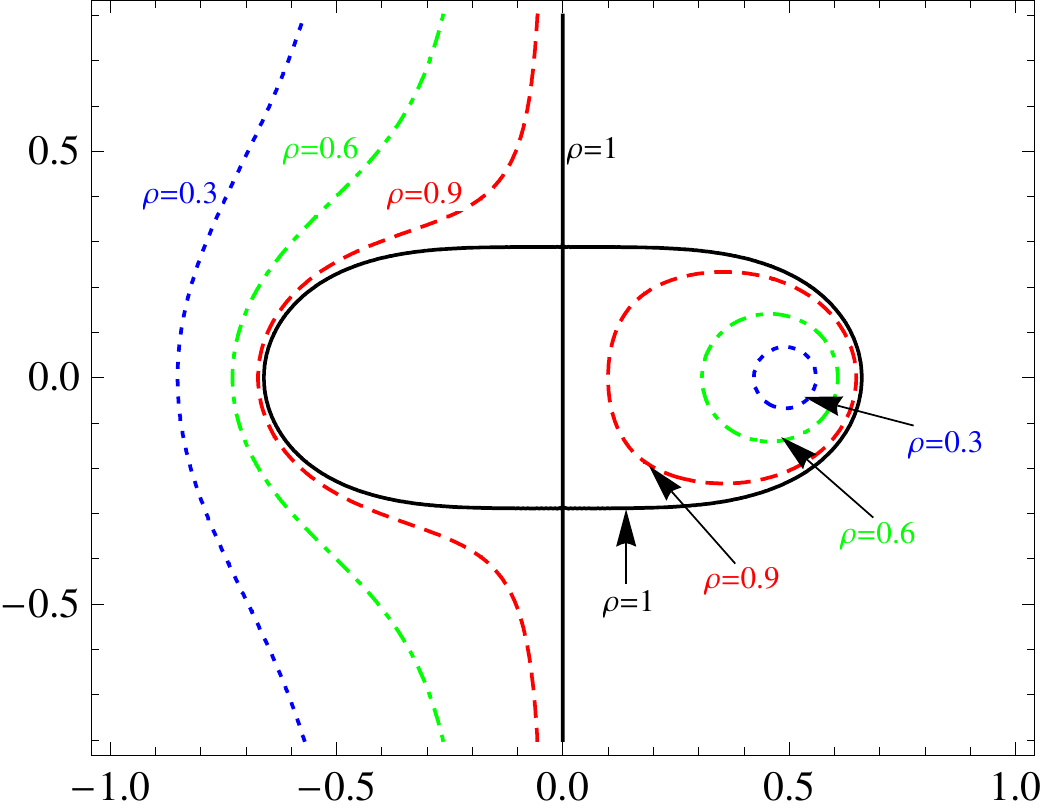}\hfill
  \includegraphics[width=0.3\textwidth]{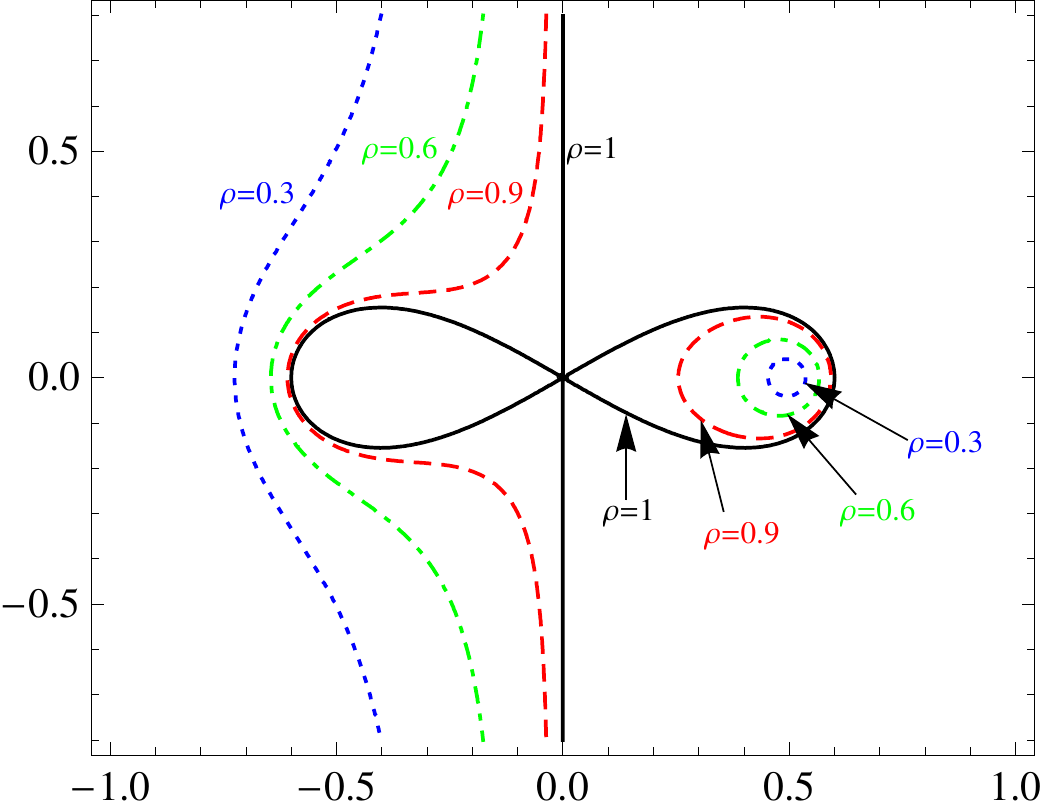}\hfill
  \includegraphics[width=0.3\textwidth]{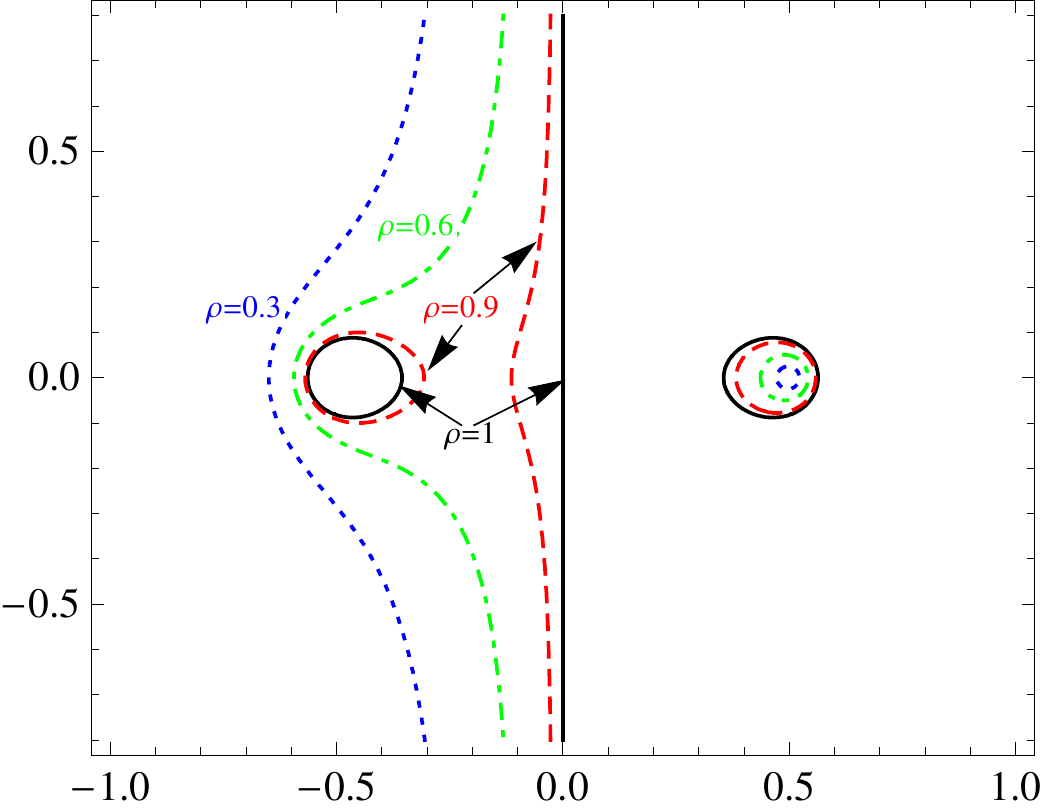}
  \caption{Contours of solutions of equation~({\protect \ref{SaddleU}}) in
    the complex-$U$ plane at $t=3$ (left), $t=4$ (middle), and $t=5$
    (right) for $\rho=1$ (black, solid), $\rho=0.9$ (red, dashed),
    $\rho=0.6$ (green, dot-dashed), and $\rho=0.3$ (blue, dotted).  In
    the figures (but not in the analysis) we have taken $|v|=1$ for
    simplicity.}
  \label{figSaddle}
\end{figure}

We now turn to the integral representation \eqref{Rint} to take the first steps in
a $1/N$ expansion of $\rho^{\tr}_N (\theta,t)$ (cf. \cite{Lohmayer:2009aw}).  
Since this integral representation was derived for $|v|<1$, we set
$v=e^{i \theta-\epsilon}$ with $|\theta|\leq\pi$, $\epsilon>0$, and
take the limit $\epsilon\to 0$ at the end.  We write~(\ref{Rint}) as
\begin{align}
  \label{RintLog}
  \bar R(v)&=-\frac{N^2}{t}e^{-\frac t2}\int\!\!\int_{-\infty}^\infty
  \frac{dxdy}{2\pi}\int_0^1d\rho\, e^{-\frac
    N{2t}\left(x^2+y^2\right)+\frac1{2t}(x+iy)^2-\frac12(x-iy)}\cr
  &\quad\times e^{(N-1)\log\left(1-v\rho
      e^{-x-t/2}\right)-(N+1)\log\left(1-v\rho e^{iy-t/2}\right)}\,.
\end{align}
At large $N$, the integrals over $x$ and $y$ decouple at leading order
and can be done independently by saddle-point approximations.  Let us
start with the integral over $y$ since it is conceptually simpler. The
$y$-dependent coefficient of the term in the exponent in
equation~(\ref{RintLog}) that is proportional to $-N$ is
\begin{equation}
  \bar f(y)=\frac1{2t} y^2+\log\left[1-v\rho e^{iy-\frac t2}\right]\,.
\end{equation}
Substituting $y=u-it/2=it(U-1/2)$ (with $u=itU$ in analogy to
section~\ref{Sec:sym}) results in exactly the same integrand that
was already considered in section~\ref{Sec:sym}, with the
replacements $T\to t$ and $z\to1/v\rho$ (with $|v\rho|<1$) and with an
integration over $u$ that is now along the line from $-\infty+it/2$ to
$+\infty+it/2$.  Since there are no singularities between this line
and the real-$u$ axis we can change the integration path to be along
the real-$u$ (or imaginary-$U$) axis.  Now everything goes through as
in section~\ref{Sec:sym}.  The saddle-point equation reads
\begin{equation}
  \label{SaddleU}
  e^{-tU}\frac{U+1/2}{U-1/2}=\frac1{v\rho}\,,
\end{equation}
which is equivalent to \eqref{contour}.  In Fig.~\ref{figSaddle} we
show the contours in the complex-$U$ plane on which the solutions of
the saddle-point equation have to lie. (For sufficiently small $\rho$
we now encounter the case mentioned in section~\ref{sec:spa} where for
$t>4$ the closed contour in the left half-plane is missing.)  The
relevant saddle point, which we denote by $y_0(\theta,t,\rho)$, is
again on the closed contour in the right half-plane.  For decreasing
$\rho$ this contour contracts, but this makes no difference to our
analysis.  The result for the $y$-integral is given by an expression
similar to \eqref{eq:spgauss}.

\begin{figure}
  \centerline{\includegraphics[width=0.5\textwidth]{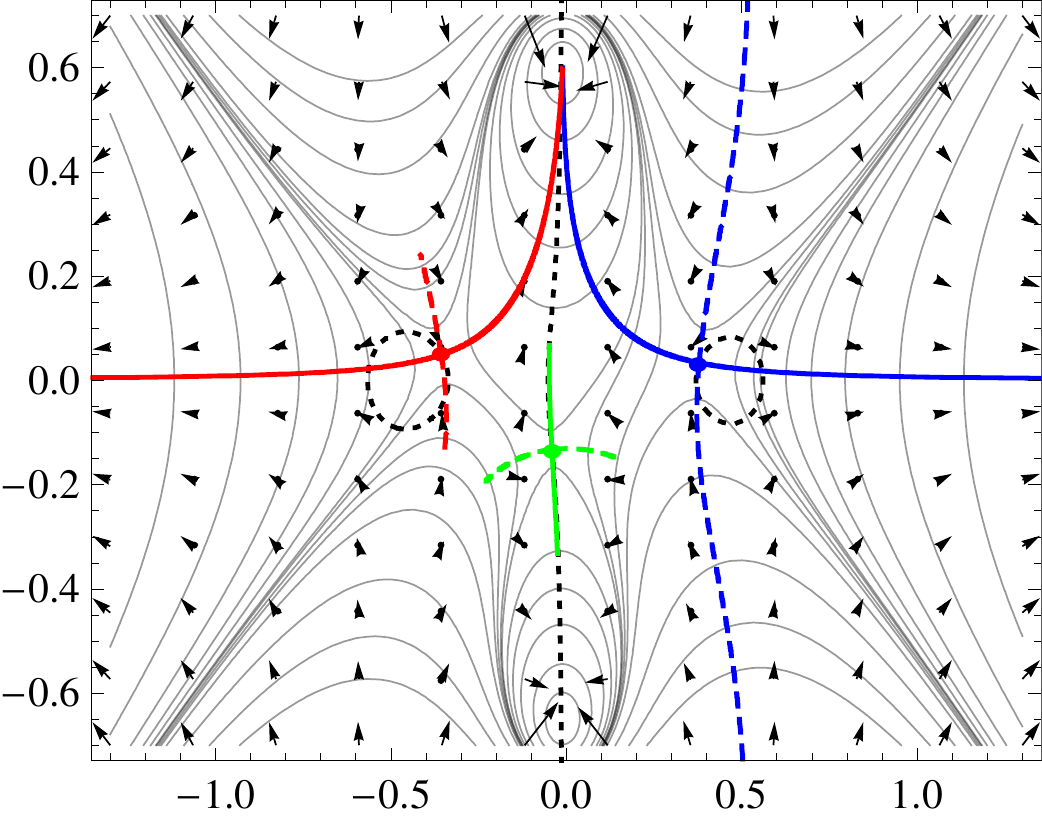}}
  \caption{Example for the location of the saddle points and the
    deformation of the integration path in the complex-$U$ plane for
    $t=5$ and $\rho=0.95$.  The dashed black curves (two closed, one
    open) are the curves on which all saddle points have to lie,
    cf.~\eqref{eq:uiur}.  In this example $\theta=3.0$.  On each of
    the closed curves there is one saddle point (red dot and blue
    dot), and on the open curve there are infinitely many saddle
    points, but only one of them in the region shown in the plot
    (green dot).  The thin solid lines are lines of constant $\re
    \tilde f(x)$ and $\re \bar f(y)$.  The arrows point in the
    direction of increasing $\re \tilde f(x)$ or decreasing $\re \bar
    f(y)$.  The dashed blue curve is the integration path for the
    $y$-integral along the direction of steepest descent.  The solid
    red-blue curve is the integration path for the $x$-integral along
    the direction of steepest descent.}
  \label{fig:profile}
\end{figure}

We now turn to the integral over $x$.  The $x$-dependent coefficient
of the term in the exponent in equation~(\ref{RintLog}) that is
proportional to $-N$ is
\begin{equation}
  \tilde f(x)=\frac1{2t} x^2-\log\left[1-v\rho
    e^{-x-t/2}\right]=-\bar f(ix)\,. 
\end{equation}
Substituting $x=-iu-t/2=t(U-1/2)$ (with $u=itU$) again leads to the
integral considered in section~\ref{Sec:sym} and the
saddle-point equation \eqref{SaddleU}, except that the integration is
now along the real-$U$ axis.  The positions of the saddle points of
the $x$-integral are obtained by rotating the saddles of the
$y$-integral by $-\pi/2$ in the complex-$U$ plane, i.e., $x_s=-iy_s$.
At a saddle point we have
\begin{equation}
  \label{fppx0}
  \tilde f''(x_s)=\frac 1t+\frac{x_s}t\left(1+\frac
    {x_s}t\right)=\bar f''(y_s) \,,
\end{equation}
and therefore the directions of steepest descent through a saddle
$y_s$ and the corresponding saddle $x_s=-iy_s$ are identical (no
rotation).  By analyzing the directions along which the phase of the
integrand is constant, we find that the integration contour can always
be deformed to go through the (single) saddle point in the right
half-plane in the direction of steepest descent.  Depending on the
parameters $\rho$, $v$, and $t$, there is either one or no additional
saddle point on the contour(s) in the left half-plane through which we
can also go in the direction of steepest descent.  If there is such an
additional saddle point, we find that its contribution to the integral
is always exponentially suppressed in $N$ compared to the saddle point
in the right half-plane and can therefore be dropped from the
saddle-point analysis.  In addition, there are infinitely many more
saddle points on the open contour in the left half-plane.  However, we
cannot deform the integration path to go through these points in the
direction of steepest descent and therefore do not need to include
them. An example for the location of the saddle points and the
deformation of the integration path is given in
Fig.~\ref{fig:profile}.  To summarize, the $x$-integral can be
approximated by the contribution of the single saddle point in the
right half-plane, which again leads to an expression similar to
\eqref{eq:spgauss}.

Combining the saddle-point approximations for the integrals over $x$
and $y$, we find that, up to exponentially small corrections in $N$,
the integral in equation~(\ref{RintLog}) is given by
\begin{equation}
  \bar R(v)=-\frac{N^2}{t}e^{-t/2}\int_0^1d\rho\,
  \frac1{2\pi}\left(\frac{2\pi}{N\tilde f''(x_0)}\right)
  \frac1{(1-v\rho e^{-x_0-t/2})^2}\,e^{-x_0}\,, 
\end{equation}
where $x_0=x_0(\theta,t,\rho)$ is the dominating saddle point of the
$x$-integral. $x_0$ is a solution of the saddle-point equation
obtained by differentiating $\tilde f(x)$, which can be written as
\begin{equation}
  \label{Saddlex0}
  v\rho e^{-x_0-t/2}=\frac{x_0}{x_0+t}
\end{equation}
and leads to
\begin{equation}
  \left(1-v\rho e^{-x_0-\frac t2}\right)^2=\left(\frac t{t+x_0}\right)^2\,.
\end{equation}
With~(\ref{fppx0}) we obtain
\begin{equation}
  \tilde f^{\prime\prime}(x_0)\left(1-v\rho e^{-x_0-\frac
      t2}\right)^2=\frac{t+x_0\left(t+x_0\right)}{\left(t+x_0\right)^2} 
\end{equation}
and
\begin{equation}
  \bar R(v)=-\frac Nt e^{-\frac t2}\int_0^1d\rho\,
  \frac{\left(t+x_0\right)^2}{t+x_0\left(t+x_0\right)}\,e^{-x_0}\,.  
\end{equation}
Differentiating equation~(\ref{Saddlex0}) with respect to $\rho$ leads to
\begin{align}
  \frac{\partial x_0}{\partial\rho}&=\frac1\rho
  \frac{x_0\left(t+x_0\right)}{t+x_0\left(t+x_0\right)}
  =ve^{-x_0-t/2}\frac{\left(t+x_0\right)^2}{t+x_0\left(t+x_0\right)}\,,
\end{align}
which yields
\begin{equation}
  \bar R(v)=-\frac{N}{tv}\int_0^1d\rho\, \frac{\partial x_0}{\partial \rho}
  =-\frac{N}{tv}\left[x_0(\theta,t,\rho=1)-x_0(\theta,t,\rho=0)\right]\,.
\end{equation}
We know from \eqref{Saddlex0} that $x_0(\theta,t,\rho=0)=0$. If we
parametrize $x_0(\theta,t,\rho=1)=\lambda(\theta,t) t$, where
$\lambda(\theta,t)$ has to solve
\begin{equation}
  \label{SaddleLambda}
  \lambda=\frac{1}{\frac1v e^{t\left(\lambda+1/2\right)}-1}\,,
\end{equation}
and take the limit $\epsilon\to0^+$, we end up with
\begin{equation}
\bar R(v)=-\frac{N\lambda(\theta,t)}v\,,\qquad v=e^{i\theta}\,.
\end{equation}
Here we need to keep in mind that we have to pick the solution of
equation~(\ref{SaddleLambda}) which corresponds to the dominating saddle
point $x_0$ of the $x$-integral for $|v\rho|<1$.

Using \eqref{eq:rho} and 
\begin{align}
G(z)=\frac1z-\frac1{z^2N}\bar R\left(z^{-1}\right)
\end{align}
 we obtain
\begin{align}
  \lim_{N\to\infty}\rho_N^\tr(\theta,t)=1+2\re\lambda(\theta,t)\,,
\end{align}
which is equal to $\rho_\infty(\theta,t)$
\cite{Durhuus:1980nb,Janik:2004tw}.  Keeping higher orders in the
saddle-point approximation (as explained in section~\ref{sec:spa1N}),
we can compute the asymptotic expansion of $\rho_N^\tr(\theta,t)$ in
powers of $1/N$.

\section{Conclusions}
The probability distribution of Wilson loops in $\SU(N)$ YM in two Euclidean dimensions can be written as a sum over irreducible representations (where only dimension, second-order Casimir, and character of $W$ enter). This allows for the derivation of integral representations for different density functions (including the true eigenvalue density), which have the same infinite-$N$ limit. These integral representations, where $N$ enters only as a parameter, are exact for any finite $N$. Results at infinite $N$ can be obtained by saddle-point approximations. Next-order terms in an expansion in $1/N$ give reasonable results in the interval where $\rho_\infty(\theta)>0$. More work is needed to get finite $N$ effects where $\rho_\infty(\theta)=0$ and corrections are exponentially suppressed in $N$.

\section{Acknowledgments}

I would like to thank Herbert Neuberger and Tilo Wettig. The presented work has been done in close collaboration with them.
I acknowledge support by BayEFG and the DOE under grant number
DE-FG02-01ER41165 at Rutgers University.

\bibliographystyle{unsrt}
\bibliography{ZakopaneLohmayer}

\end{document}